\documentstyle[twocolumn,aps,psfig]{revtex}
\begin{document}
\draft
\twocolumn[\hsize\textwidth\columnwidth\hsize\csname @twocolumnfalse\endcsname
%
%
%

\title{Pseudogap Formation in Models for Manganites}

\author{ Adriana Moreo, Seiji Yunoki, and Elbio Dagotto}

\address{National High Magnetic Field Lab and Department of Physics,
Florida State University, Tallahassee, FL 32306}

\date{\today}
\maketitle

\begin{abstract}

The density-of-states (DOS) and one-particle spectral function $\rm A({\bf k},
\omega)$ of the one- and two-orbital models for manganites, the latter with
Jahn-Teller phonons, are evaluated using Monte Carlo techniques. 
Unexpectedly robust pseudogap (PG) features were found
at low- and intermediate-temperatures, particularly at or near regimes
where phase-separation occurs as $\rm T$$\rightarrow$0. The PG follows 
the chemical potential and it is caused by the formation of
ferromagnetic metallic clusters in an insulating
background. It is argued that PG formation should be generic of
mixed-phase regimes.
The results are in good agreement with recent photoemission
experiments for $\rm La_{1.2} Sr_{1.8} Mn_2 O_7$.

\end{abstract}
\vskip2pc]
\narrowtext

%
%

Manganese oxides have been intensively studied in recent years due to their
colossal magnetoresistance (CMR) effect\cite{jin}, where small changes in
magnetic field induce large changes in their resistivity 
$\rho_{dc}$. The CMR behavior may
originate in the nontrivial interplay of the charge, spin, orbital,
and lattice 
degrees of freedom in these materials.
Among the theories proposed for the
CMR effect are those based on the recently observed
intrinsic tendencies of models for
manganites toward mixed-phase formation\cite{yunoki,yunoki2}.
In this context, clusters of one phase embedded
into the other, typically involving 
ferromagnetic (FM) metallic and FM or antiferromagnetic (AF)
insulating phases,
 are believed to
substantially increase $\rho_{dc}$ and the
compressibility $\kappa$ of the system\cite{science}. The effect 
occurs even in stable metallic regimes
at low-temperatures, as long as the density is close to those where
phase-separation occurs\cite{science}. 
A variety of experimental results are in qualitative
agreement with these ideas and evidence is accumulating that
manganites have important
microscopic inhomogeneities, particularly in regimes of
relevance for the CMR effect\cite{science,exper}.

Recently,  high-energy resolution
angle-resolved photoemission (ARPES) measurements for 
the bilayer compound $\rm La_{1.2} Sr_{1.8} Mn_2 O_7$ (hole
density x=0.4) have been reported\cite{dessau}. 
The most remarkable result
was the existence of a $pseudogap$ at the
chemical potential $\rm \mu$, 
substantially
larger than the analog feature reported for the  cuprates.
The main purpose of this paper is to analyze whether similar PG
characteristics appear in the DOS of manganite models when analyzed 
using unbiased computational many-body methods\cite{yunoki}.
The results are surprisingly robust, showing that indeed a prominent 
pseudogap exists in the DOS, as long as the regime explored has
mixed-phase characteristics.
The conclusions are general and they should apply to other compounds
with similar microscopic phase-separation tendencies as well.

The one-orbital Hamiltonian is given by
$\rm H= H_{DE} + H_{AF}$, with the Double-Exchange (DE) term defined as~\cite{zener}
$$
\rm H_{DE} = -t \sum_{\langle {\bf i j} \rangle \sigma} (c^\dagger_{{\bf i}\sigma} c_{{\bf
j}\sigma} + h.c.) - J_H \sum_{{\bf i}\alpha \beta}
{ {{\bf S}_{\bf i}}
\cdot{ c^{\dagger}_{{\bf i}\alpha} {\bf \sigma}_{\alpha \beta} c_{{\bf
i}\beta} } },
\eqno(1)
$$
\noindent where $\rm c^{\dagger}_{{\bf i}\sigma}$ creates
an $\rm e_g$-fermion at site ${\bf i}$
with spin $\sigma$, and $\rm {\bf S}_{\bf i}$ is the total 
$\rm t_{2g}$-spin, assumed 
localized and classical ($\rm |{\bf S}_{\bf i}|$=1).
$\rm J_H$$>$0 is the Hund coupling,
and the rest of the notation is standard. 
$\rm H_{AF} = J' \sum_{\bf \langle ij \rangle} 
{{{\bf S}_{\bf i}}\cdot{{\bf S}_{\bf j}}} $ is a direct AF
coupling between  localized spins. 
The two-orbitals Hamiltonian is
$\rm H_{KJT} = H_K + H_{JT} + H_{AF}$. The first term containing the
kinetic energy and Hund coupling is 
$$
\rm H_K = -\sum_{{\bf i} {\bf u} \sigma a b} t^u_{ab}
(c^\dagger_{{\bf i} a \sigma} c_{{\bf i+u} b \sigma} + h.c.)
- J_H \sum_{{\bf i}} { {{\bf S}_{\bf i}}\cdot{ {\bf s}_{\bf i}
} },
\eqno{({\rm 2a})}
$$
\noindent 
where $\rm a,b$=1,2 are the two $\rm e_g$-orbitals, ${\bf u}$ are unit vectors
along the main axes, u labels those axes,
$\rm {\bf s}_{\bf i} = \sum_{a \alpha \beta}
c^\dagger_{{\bf i}a \alpha} {\bf \sigma}_{\alpha \beta} c_{{\bf i}a \beta}$
is the spin of the mobile fermions,     
and the rest of the notation is standard\cite{hopping}. 
In this paper, the unit of energy will be
 $\rm t$=1 in Eq.(1),
$\rm t^x_{11}=1$ in Eq.(2a), and $\rm J'$
will be fixed to 0.05\cite{comm1}.
The coupling with JT-phonons
($\rm Q^{11}_{\bf i}$=$\rm -Q^{22}_{\bf i}$=$\rm 
Q^{(3)}_{\bf i}$, and
$\rm Q^{12}_{\bf i}$=$\rm Q^{21}_{\bf i}$=$\rm Q^{(2)}_{\bf i}$) is~\cite{millis}
$$
\rm H_{JT} = \lambda \sum_{{\bf i} a b \sigma} c^{\dagger}_{{\bf i} a
\sigma}
Q^{ab}_{\bf i} c_{{\bf i} b
\sigma}
+ {{1}\over{2}} \sum_{\bf i} ( {Q^{(2)}}^2_{\bf i} + {Q^{(3)}}^2_{\bf i}).
\eqno{({\rm 2b})}
$$
\noindent 
Classical phonons 
are assumed for simplicity.
The $\rm e_g$-fermionic density $\rm \langle n \rangle$ is adjusted with
$\mu$, for both models Eqs.(1,2) (the hole-density
is $\rm x=1-\langle n \rangle$).
To study their properties,  a Monte Carlo
(MC) algorithm for the classical spins and phonons  and an
exact diagonalization of a
one-electron problem for a given spin-phonon background were used. The method 
allows dynamical calculations in real-time without 
uncontrolled analytical continuations, and it
has been described in detail in previous publications\cite{yunoki}.
The DOS $\rm N(\omega)$ was obtained adding the
one-particle spectral functions $\rm A({\bf k},\omega)$, and
it is directly related with the DOS of photoemission experiments.

Fig.1a contains $\rm N(\omega)$
of the one-dimensional (1D) one-orbital model Eq.(1) at large Hund coupling
($\rm J_H$=$\infty$ for simplicity) and $\rm \langle n
\rangle$$ \sim$0.88 ($\rm x \sim 0.12$). In this regime, 
MC calculations of $\rm \langle n \rangle$ vs $\mu$
at zero temperature (T) similar to those reported in Refs.\cite{yunoki,yunoki2}
revealed the presence of phase-separation ($\kappa$$\rightarrow$$\infty$)
for $\rm 0.77$$<$$\rm \langle n \rangle$$<$1.0. As a consequence,
the finite-T state investigated in
Fig.1a presents a large
compressibility.
Size effects are small, and the results
are expected to be representative of the bulk limit.
The most noticeable feature of Fig.1a is the deep
minimum at $\omega$=$\mu$, which clearly develops 
as T is reduced. 
Similar behavior has been observed
for other parameters ($\rm J_H$, $\rm J'$, T and $\rm
\langle n \rangle$) investigated here, as long as $\kappa$ is
large. Fig.1b shows two-dimensional (2D)
results, still for the
one-orbital model. Here phase-separation at T=0 was found in 
the interval $\rm 0.86$$<$$\rm \langle n \rangle$$<$$1.0$.
Fixing T to a low value, and varying
$\rm \langle n \rangle$, PG behavior is once again
observed where T=0 phase-separation occurs.
In Figs.1a-b the PG
is always centered at $\mu$, i.e. the DOS is 
not rigid but adjusts with $\rm \langle n \rangle$ and T.
Also note that at densities with marginally
 stable T=0 ground-states,
PG remnants are observed (insets of Figs.1a-b),
as long as $\kappa$ remains large. This
occurs in a narrow $\rm \langle n \rangle$-range near 
phase-separated regimes.

\begin{figure}[htbp]
\vspace{-0.7cm}
\centerline{\psfig{figure=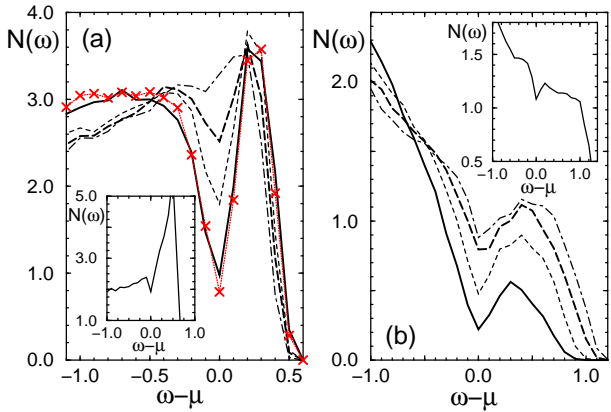,width=8.5cm,angle=-90}}
\vspace{0.2cm}
\caption{
Pseudogap formation in 
the one-orbital model, as a function of $\rm \langle n \rangle$ and T.
(a) Density of states $\rm N(\omega)$ vs $\omega$-$\mu$
at $\rm J_H$=$ \infty$ and 1D.
The overall density is $\rm \langle n \rangle$$\sim$$0.88$ 
($\rm x \sim 0.12$). 
Starting from the top at $\omega$-$\mu$=0, the temperatures $\rm T$ of
the first four lines
are 1/10, 1/17, 1/25, 1/40, and the lattice size is L=30. The dotted
line with crosses was obtained at T=1/40, but with L=100 showing
the absence of strong size effects. The inset contains the DOS
for L=100, T=1/40, and $\rm \langle n
\rangle$$\sim$0.77, a marginally 
stable density as T$\rm  \rightarrow$0; (b) 
$\rm N(\omega)$ vs $\omega$-$\mu$
in 2D at $\rm J_H$=$\infty$, 
using a $10 \times 10$ cluster and T=1/30. The four lines
from the top at $\omega$-$\mu$=0 correspond to
$\rm \langle n \rangle $ $\sim$0.90, 0.92, 0.94, and 0.97. The
inset has results for $\rm \langle n \rangle$=$0.86$, a marginally
stable density at T=0.
}
\vspace{0.2cm}
\end{figure}

Pseudogap features appear also in the two-orbital model with
JT-phonons, again
as long as $\kappa$ is robust. For instance, working at large
$\rm J_H$ and $\lambda$, Fig.2a shows results at
$\rm \langle n \rangle$$\sim$0.7, 
varying T. For the parameters of Fig.2a,
MC studies showed that T=0
phase-separation occurs in the interval
$\rm 0.5$$<$$\rm \langle n \rangle$$<$1.0 between spin FM-phases that
differ in the orbital-order pattern (uniform vs staggered)\cite{yunoki2}.
Fig.2a
illustrates the PG development as T is reduced, similarly as 
for the one-orbital model. An analogous phenomenon  also occurs
at low-density (Fig.2b) where T=0
phase-separation between spin FM- and AF-phases
 was numerically observed for
$\rm 0.0$$<$$\rm \langle n \rangle$$<$0.3\cite{yunoki2}.
At low $\rm \langle n \rangle$, results without electron-phonon
coupling (Fig.3a)
are similar to those 
with strong lattice coupling $\lambda = 1.5$.
Vestiges of the PG also appear at $\rm \langle n
\rangle \sim 0.4$ (Fig.2c), which has a stable T=0 ground-state based on
previous studies\cite{yunoki2}.
A similar situation occurs at $\lambda=1$ and $\rm \langle n \rangle
\sim 0.7$, near the phase-separation boundary (Fig.3b).

\begin{figure}[htbp]
\vspace{-0.7cm}
\centerline{\psfig{figure=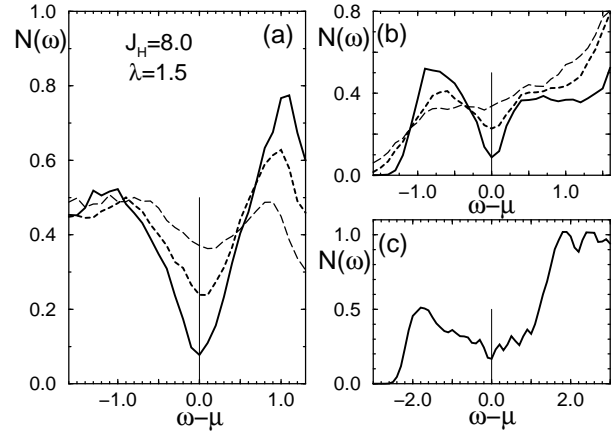,width=8.5cm,angle=-90}}
\vspace{0.2cm}
\caption{
Pseudogap behavior
of the 1D two-orbitals model at large electron-phonon
coupling, as a function of $\rm \langle n \rangle$ and T.
(a) $\rm N(\omega)$ vs $\omega$-$\mu$ with
$\rm J_H$=$8$, $\lambda=1.5$, L=20, and
$\rm \langle n \rangle \sim 0.7$ ($\rm x \sim 0.3$). 
The thin-dashed, thick-dashed, and solid
lines correspond to T=1/5, 1/10 and 1/20, respectively;
(b) Same as (a) but at $\rm J_H=\infty$, L=22 and 
$\rm \langle n \rangle \sim 0.2$. The thin-dashed, thick-dashed, and solid
lines correspond to T=1/5, 1/10 and 1/25, respectively;
(c) Same as (a) but for T=1/15 and 
$\rm \langle n \rangle \sim 0.4$, a stable density as 
$\rm T \rightarrow 0$.
}
\vspace{0.2cm}
\end{figure}

To investigate the origin of the PG, in Fig.3c 
the charge-charge correlation in momentum-space $\rm N(q)$ is shown 
in the region of interest. At low hole-density,
$\rm N(q)$ is enhanced
 at small-q in the PG regime,
indicating the existence of extended charged structures. Similar
conclusions are reached by noticing that the average FM-cluster size
increases as T is reduced (Fig.3d).
The overall
study of the data accumulated here lead us to
conclude that the PG formation is
correlated with large
charge fluctuations, 
which occur both with and without phonons in regions
that phase-separate as T$\rightarrow$0, and in their vicinity.

Although the relation between PG formation and mixed-phase tendencies is
clear,  the sharpness of the PG feature and its
location following $\mu$  as $\rm \langle n \rangle$
is varied requires additional analysis.
For this purpose, several
 $\rm t_{2g}$-spin MC-configurations (snapshots) of the 1D one-orbital model 
(Fig.1a) were individually studied (the reasoning and results described
 below apply with minor
changes to the two-orbital model with JT-phonons and in 
any dimension).
The analysis of the one-electron energies
revealed that most of the MC
spin configurations produce the
PG feature, and, thus,
a single MC-generated snapshot
should be enough to understand the
phenomenon. Visual analysis indicated that
the configurations with PG contain FM clusters of an intermediate size
between 2 to 6 lattice spacings, immersed in a background
with AF-correlations (short-ranged 
since here T is comparable to $\rm J'$).
To understand why this spin arrangement produces a PG, 
consider for simplicity the low-$\rm \langle n
\rangle$ limit. Since at $\rm J_H=\infty$, the high and low density limits are
identical, while at finite but large $\rm J_H$ they are very similar,
our analysis applies to $\rm \langle n \rangle$ close to 1 simply changing
electrons by holes.
A cartoon-like representation of a typical MC $\rm t_{2g}$-spin snapshot
is in Fig.4a\cite{2D}. The electrons are correlated with the classical
spins such that  they mostly populate the FM regions (Fig.4b), 
and, thus, aligned $\rm t_{2g}$-spins provide to the electrons
 an attractive effective 
 potential
(Fig.4c). When extra electrons are added, FM clusters (with one or more
electrons) are created,
and new (occupied) levels appear below $\mu$.
Then, the one-particle spectrum at $\omega < \mu$
for one MC classical spin configuration
 contains energies corresponding to 
 quasi-localized electrons in each FM-well, with
a finite bandwidth  (``cluster'' band)
caused by the different shapes of those wells, plus 
tunneling among them. 
The AF-regions contribute to $\omega > \mu$.
As a consequence, the resulting DOS (Fig.4d) contains a
pseudogap at $\rm \omega = \mu$.

\begin{figure}[htbp]
\vspace{-0.7cm}
\centerline{\psfig{figure=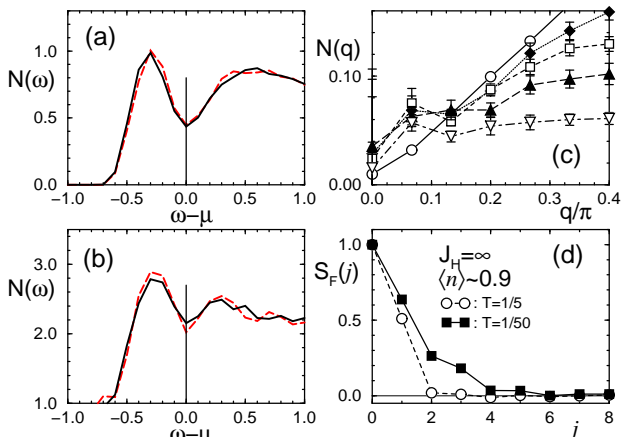,width=8.5cm,angle=-90}}
\vspace{0.2cm}
\caption{ 
(a) $\rm N(\omega)$ vs $\omega$-$\mu$ for the 1D two-orbitals model 
without phonons ($\lambda =0$) at low-density 
$\rm \langle n \rangle \sim 0.1$, $\rm J_H=\infty$, and T=1/20. 
Shown are sizes L=40 and
22, denoted by solid and dashed lines, respectively; (b) Same as (a) but
at $\lambda = 1.0$, T=1/25 and $\rm \langle n \rangle \sim 0.7$. 
This is at the boundary of the T=0 phase-separated region [3]. 
The solid and dashed lines correspond to L=40 and 50,
respectively; (c) N(q) for the 1D one-orbital model with 
$\rm J_H$=$\infty$, T=1/30, and L=30. Open circles, full diamonds, open
squares, full triangles, and open triangles, denote 
$\rm \langle n \rangle=0.75$, 0.84, 0.87, 0.90, and 0.94, respectively;
(d) Size of typical FM clusters for the 1D 
one-orbital model. $\rm S_F(j)$ (normalized to $\rm S_F(0)$) is the classical 
spin-spin correlation 
$\rm \langle {\bf S}_i \cdot {\bf S}_{i+j} \rangle$
($\rm j>i$) measured only when $\rm S_F(1)$ is positive as a way to 
isolate FM clusters. 
}
\vspace{0.2cm}
\end{figure}

The present study also included the individual one-particle spectral
functions $\rm A({\bf k}, \omega)$.
Figs.5a-b show the
2D one-orbital model at large $\rm J_H$, low-T, and low hole-density.
The data displays the PG feature, which
appears at $all$
values of the momenta which have appreciable weight at $\omega$=$\mu$. Similar
behavior is obtained for the two-orbitals model at large $\lambda$
(Fig.5c). These results are
compatible with the explanation of PG behavior based on
mixed-phase tendencies, since the clusters formed in such a state are
not distributed forming a regular pattern. 
Then, there is no preferred momentum for the PG to occur, unlike in a
charge-density-wave state. The results of Fig.5 are 
in excellent qualitative agreement with ARPES measurements 
for $\rm La_{1.2} Sr_{1.8} Mn_2 O_7$\cite{dessau}, where
the PG was observed all along the ``Fermi surface''. 
Note also the dispersive character of the results of Fig.5, 
in qualitative 
agreement with the ARPES data\cite{dessau}. 
They originate mainly from the
AF-regions that induce an effective DE-hopping
$\rm t_{eff}$=$\rm t\langle cos(\theta/2)\rangle$ which for the
temperatures in Figs.5a-b is approximately $\rm t/2$.

\begin{figure}[htbp]
\vspace{0.3cm}
\centerline{\psfig{figure=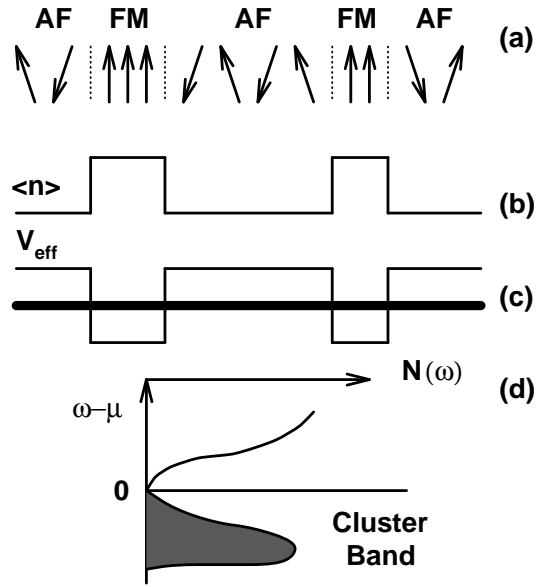,width=7.0cm,angle=-0}}
\vspace{0.2cm}
\caption{
Schematic explanation of the PG formation in the low-$\rm \langle
n \rangle$ limit and for the one-orbital model. In (a) a typical MC
$\rm t_{2g}$-spin configuration is sketched.
In (b), the corresponding $\rm e_g$-density is shown, with electrons
mostly located in the FM regions. In (c), the effective potential felt by the
electrons is presented. A (populated)
``cluster band'' is formed (thick line). In (d) the resulting DOS is shown.
}
\vspace{0.1cm}
\end{figure}

Other unexpected effects in ARPES data are also reproduced in
Fig.5. For instance, the width of the peaks found in Ref.\cite{dessau} were
as large as 0.5-1.0 eV, similar to  the widths
in Fig.5 which are of order 1-2t, with t 
estimated to be 0.2-0.4 eV\cite{yunoki}.
The large width is caused by the disordering influence 
of localized spins and phonons,
particularly within the AF-clusters which have 
only partially ordered spins at the T's analyzed here. The
peaks do not 
sharpen as $\mu$ is reached, 
as in experiments\cite{dessau}, and the concept of
quasiparticle seems not applicable in the models studied here.

Motivated by the similarities of our results with those of ARPES
studies for the
bilayer manganites\cite{dessau},
here it is conjectured that
the x=0.4 state of $\rm La_{1.2} Sr_{1.8} Mn_2 O_7$ 
should present microscopic inhomogeneities, both above and 
below $\rm T_C$. This is compatible with
a recently reported neutron-scattering-based
 phase diagram\cite{kubota}, 
where coexistence of FM (metal) and AF
(insulator) features were observed at x=0.4 and low-T.
Our prediction is also compatible with other neutron scattering 
results\cite{perring} 
that reported mixed AF/FM characteristics and
short-range charge ordering due to cluster formation, for the
same compound above $\rm T_C$.

\begin{figure}[htbp]
\vspace{-0.7cm}
\centerline{\psfig{figure=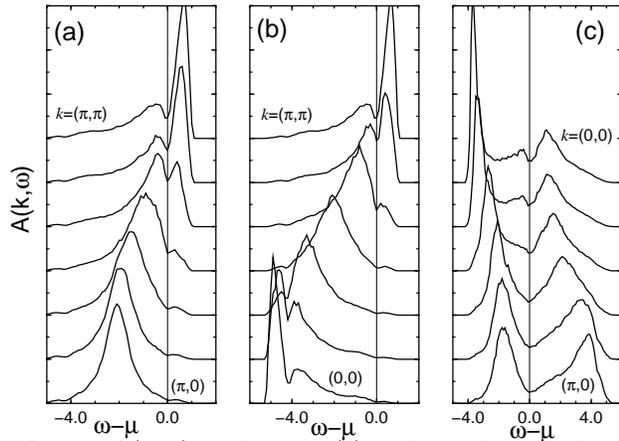,width=8.5cm,angle=-90}}
\vspace{0.2cm}
\caption{
$\rm A({\bf k}, \omega)$ at $\rm
J_H$=$\infty$.
(a) is for a 2D one-orbital case with 
T=1/30, $\rm \langle n \rangle \sim 0.92$ ($\rm x \sim 0.08$), and on 
a $12 \times 12$ cluster with periodic boundary conditions 
(PBC). Shown are results along
$(0,0)$ to $(\pi,0)$; (b) Same as (a) but along $(0,0)$ to
$(\pi,\pi)$; (c) Results for the 2D two-orbitals model with T=1/10,
$\lambda$=$1.5$, $\rm \langle n \rangle \sim 0.7$ ($\rm x \sim 0.3$), 
on a $10 \times 10$
cluster, along $(0,0)$ to $(\pi,0)$, with PBC.
}
\vspace{0.2cm}
\end{figure}

The general character of our explanation of the computational results
suggests that 
PG behavior should be present in other
compounds if they are in mixed-phase regimes. 
Previous studies\cite{science,exper} 
suggest that this may occur in the small x region 
of $\rm La_{1-x} Sr_x Mn O_3$ at low- and intermediate-T, 
as well as within the FM metallic phase at low-T close
to the metal-insulator transition where $\kappa$
is expected to be large\cite{science}. 
On the other hand, at x$\sim$0.3 this compound
behaves more as a regular metal and it should not have a PG. In general,
where $\rm T_C$ is the largest is where PG tendencies
should be the weakest, correlated with weak CMR effects.
In $\rm La_{1-x} Ca_x Mn O_3$ at low-T, the interesting region is again
small x, but also $\rm x \sim 0.5$ where phase-separation 
has been observed\cite{exper}. If the insulating character of
this compound
above $\rm T_C$ is caused by microscopic inhomogeneities, PG
features should also appear in a wide range of compositions at intermediate-T.

Summarizing, simple models for manganites
studied computationally
have a DOS with robust pseudogap features
in regions of parameter space with large compressibility, particularly low
hole-doping. 
$\rm N(\mu)$ is drastically reduced as T is lowered
in this regime. 
The PG was found for all momenta along the ``Fermi surface'', in agreement
with experiments. 
The effect is caused by the formation of FM metallic clusters in an
insulating background, precursors of the T=0 phase-separation tendencies
which occurs both with and without JT-phonons.
Mixed-phase tendencies
are crucial for PG formation
and its associated $\rho_{dc}$ increase. This conclusion
seems independent of the details of the models, and may occur 
even if cluster formation is caused by 
1/r-Coulomb interactions
on phase-separated regions.
The close relation between $\rm N(\mu)$ and $\rho_{dc}$ expected
here differs drastically from theories based on Anderson localization,
or in a simple
hopping reduction by spin fluctuations
$\rm t$$\rightarrow$$\rm t\langle cos(\theta/2)\rangle$, where
 $\rm N(\mu)$ is 
approximately unchanged\cite{smol}.
The mechanism discussed in this paper can also be operative 
in non Jahn-Teller
compounds with tendencies to phase-separation, such as
diluted magnetic semiconductors.

Discussions with D. Dessau and M. Kubota are acknowledged.
The authors are supported in part by grant NSF-DMR-9814350.

\medskip

\vfil

%
%


\begin{references}


\bibitem{jin} Jin et al., Science {\bf 264}, 413 (1994).

\bibitem{yunoki} S. Yunoki {\it et al.}, Phys. Rev. Lett. {\bf 80}, 845
(1998); S. Yunoki and A. Moreo, Phys. Rev. B{\bf 58}, 6403 (1998).

\bibitem{yunoki2} S. Yunoki {\it et al.}, 
Phys. Rev. Lett. {\bf 81}, 5612 (1998).

\bibitem{science} A. Moreo {\it et al.}, Science {\bf 283}, 2034 (1999).

\bibitem{exper} See for instance
T. Egami, J. of Low Temp. Phys. {\bf 105}, 791 (1996);
M. Hennion {\it et al.}, Phys. Rev. Lett. {\bf 81}, 1957 (1998);
S. Mori et al., C. H. Chen, and S-W. Cheong, 
Phys. Rev. Lett. {\bf 81}, 3972 (1998).
For additional references see Ref.\cite{science}.

\bibitem{dessau} D. Dessau {\it et al.}, Phys. Rev. Lett. {\bf 81}, 192
(1998); D. Dessau and Z.X. Shen, in ``Colossal
Magnetoresistive Oxides'', Ed. Y. Tokura, Gordon and Breach 1999.

\bibitem{zener} C. Zener, Phys. Rev. {\bf 82}, 403 (1951).

\bibitem{hopping} In 1D 
$\rm t_{11}$=$\rm t_{22}$=$\rm 2 t_{12}$=$\rm  2 t_{21}$ was used.
In 2D, the sets
$\rm t^y_{11}$=$\rm 3 t^y_{22}$=$\rm \sqrt{3} t^y_{12}$=$\rm \sqrt{3} 
t^y_{21}$ and
$\rm t^x_{11}$=$\rm 3 t^x_{22}$=$\rm -\sqrt{3} t^x_{12}$=$\rm -\sqrt{3} 
t^x_{21}$ were used (S. Ishihara {\it et al.}, Phys. Rev.
B{\bf 56}, 686 (1997)).


\bibitem{comm1} This $\rm J'$ reproduces the N\'eel temperature of
 the fully doped $\rm Ca Mn O_3$ material.

\bibitem{millis} A. J. Millis, et al.,
Phys. Rev. Lett. {\bf 74}, 5144 (1995).

\bibitem{2D} In 2D
it was observed that the FM clusters have a variety of shapes including
a large percentage with elongated 1D-like characteristics.



\bibitem{kubota} M. Kubota {\it et al.}, cond-mat/9902288, and private
communications. 

\bibitem{perring} T. G. Perring et al., Phys. Rev. Lett. {\bf 78}, 
3197 (1997); L. Vasiliu-Doloc {\it et al.}, preprint.


\bibitem{smol}
$\rho_{dc}$ measurements on 
$\rm La_{0.67} Ca_{0.33} Mn O_3$ thin-films also provides evidence
against Anderson localization
(V. N. Smolyaninova {\it et al.}, cond-mat/9903238).

\end{references}
\end{document}